\documentclass[pra,aps,preprint,prabib]{revtex4}
\usepackage{graphicx,longtable,enumerate}
\usepackage{dcolumn}
\usepackage{bm}
\begin{document}
\title{Atomic excitations during the nuclear $\beta^{-}$ decay in light
       atoms}
\author{Alexei M. Frolov\dag and Mar\'{\i}a Bel\'en Ruiz\ddag}
\address{\dag \it Department of Chemistry, University of Western Ontario,
     London, Ontario, Canada N6A 3K7}
\address{\ddag \it Department of Theoretical Chemistry of the
Friedrich-Alexander-University Erlangen-N\"urnberg, Egerlandstra\ss e, D-91058
Erlangen, Germany}
\date{July 27, 2010}
\begin{abstract}
Probabilities of various final states are determined numerically for a
number of $\beta^{-}$-decaying light atoms. In our evaluations of the final
state probabilities we have used the highly accurate atomic wave functions
constructed for each few-electron atom/ion. We also discuss an experimental
possibility to observe negatively charged ions which form during the nuclear
$\beta^{+}$-decays. High order corrections to the results obtained for
$\beta^{\pm}$-decays in few-electron atoms with the use of sudden
approximation are considered.\\

PACS number(s): 36.10.Dr, 13.35.Bv and 14.60.Ef
\end{abstract}

\maketitle
\newpage

\section{Introduction}

In this study we consider atomic excitations which arise during the nuclear
$\beta^{-}$ decay in light few-electron atoms. Our main goal is to determine
numerically the corresponding final state probabilities, or, in other words,
the absolute probabilities of formation of the final system(s) in certain
bound and/or unbound states which arise after the nuclear $\beta^{-}$ decay
in light few-electron atoms. A basic theoretical analysis of atomic
excitations during the nuclear $\beta^{-}$ decay has been performed in our
earlier works \cite{FrTa} and \cite{Fr98}. In this study we will not repeat
all steps and arguments from those works. Instead, below we shall bring our
attention to some new problems which have not been solved in earlier
studies. Note only that our analysis and computations of atomic excitations
are based on the sudden approximation \cite{Mig1}, \cite{Mig2}. In turn, the
sudden approximation follows from the well known experimental fact that the
velocities of the emitted $\beta^{-}$ electrons are significantly larger
than the usual velocities of atomic electrons. In many actual cases such
velocities are close to the speed of light in vacuum, i.e. $v_{\beta}
\approx c$. It follows from here that the emitted $\beta^-$ electron leaves
the external shells of an atom for a time which is approximately
$\tau_{\beta} \approx a_0 / c = \alpha \tau_a = \alpha \hbar / (e^4 m_e)$,
where $\alpha \approx \frac{1}{137}$ is the fine structure constant, $\hbar$
is the reduced Planck constant, $m_e$ is the electron mass (at rest), $a_0$
is the Bohr radius, $c$ is the speed of light in vacuum and $\tau_a = \hbar
/ (e^4 m_e) \approx 2.418884 \cdot 10^{-17}$ $sec$ is the atomic time. For
internal atomic/electron shells one also finds that $\tau_{\beta} \ll
\tau_a$, since the passing time $\tau_{\beta}$ for $\beta^-$ electron
decreases with the radius of the electron shell.

The general equation of the $\beta^-$ decay can be written in the form
\begin{equation}
 Q \rightarrow (Q + 1)^{+} + e^{-} + \nu
\end{equation}
where $Q$ is the nuclear charge of the incident nucleus, while $e^-$ and
$\nu$ are the emitted (fast) electron and neutrino, respectively. The
emitted electron is usually very fast and its Lorentz $\gamma-$factor
($\gamma = E / m_e c^2$) is bounded between 2 and 15 - 18. In all actual
cases, the nuclear $\beta^-$ decay proceeds in many-electron atoms/ions,
rather than in bare nuclei. The arising atomic system with the nuclear
charge $(Q + 1)^{+}$ is also many-electron ion (or atom). Our main goal in
this study is to determine the final state probabilities for this newly
arising atomic system. Suppose that the incident atom was in one of its
bound states, e.g., in the $A$-state. The final ion is formed in one of its
states (bound or unbound), e.g., in the $B$-state. The aim of theoretical
analysis of nuclear $\beta^{\pm}$ decays in atomic systems is to evaluate
the corresponding transition amplitude ${\cal A}_{AB} = \mid \langle A \mid
B \rangle \mid$ and final state probability $p_{AB} = {\cal A}^2_{AB} = \mid
\langle A \mid B \rangle \mid^2$.

This problem has attracted a significant theoretical attention (see, e.g.,
\cite{FrTa}, \cite{Finb}, \cite{Skoro}, \cite{Schw}), since various
$\beta^{-}$ decaying nuclei are of great interest in various applications to
modern technology, scientific research, nuclear medicine, etc. For instance,
the $\beta^{-}$ decaying isotope ${}^{131}$I (so-called `radioiodine') is
extensively used in nuclear medicine both diagnostically and
therapeutically. Examples of its use in radiation therapy include the
treatment of thyrotoxicosis and thyroid cancer. Diagnostic tests exploit the
mechanism of absorption of iodine by the normal cells of the thyroid gland.
Iodine-131 can be used to destroy thyroid cells theraputically. Other
$\beta^{-}$ decaying isotopes of iodine are used (mainly as a radioactive
labels) in modern biology, physical and organic chemistry \cite{HaAd}.

Another well known $\beta^{-}$ decaying isotope is strontium-90. It finds
extensive use in medicine and industry, as a radioactive source for
thickness gauges and for superficial of some cancers. Controlled amounts
of ${}^{90}$Sr and can be used in treatment of bone cancer. The radioactive
decay of strontium-90 generates significant amount of heat. Strontium
fluoride of strontium-90 (${}^{90}$SrF$_2$) is widely used as a heat source
in many remote thermoelectric generators, since it is much cheaper and less
dangerous than the alternative source based on ${}^{238}$Pu. Strontium-90 is
also used as a radioactive tracer in medicine and agriculture. The isotope
${}^{90}$Sr can be found in significant amount in spent nuclear fuel and in
radioactive waste from nuclear reactors and in nuclear fallout from nuclear
tests. It is interesting to note that the fission product yield of
${}^{90}$Sr sharply depends upon the type of explosive nuclear (fission)
device. Relatively large output of ${}^{90}$Sr in the nuclear fallout is a
strong indication that the original nuclear explosive device was made from
uranium-233 (or uranium-235), rather than from plutonium-239. Advanced
nuclear explosive devices which contain substantial amounts of
${}^{245}$Cm/${}^{247}$Cm and/or ${}^{249}$Cf/${}^{251}$Cf produce
significantly smaller yields of ${}^{90}$Sr, than analogous devices made
from ${}^{239}$Pu. A brief discussion of different applications of other
$\beta^{-}$ decaying atoms can be found, e.g., in \cite{FrTa} (see also
\cite{HaAd}). Note that for ${}^{131}$I, ${}^{90}$Sr and for many other
$\beta^{-}$ decaying isotopes/atoms our knowledge about the final (or
post-decay) atomic states is far from complete, since in almost all cases we
cannot determine the final state probabilities. Currently, for some of the
$\beta^{-}$ decaying atoms we can only predict approximate probabilities to
find the final ions/atoms in their ground state(s). Analogous evaluations of
for the probability to form the first excited (bound) states and for the
total probability of electron ionization are very approximate. Probabilities
to form other excited states, including various unbound states, in the final
atomic systems have never been evaluated (even approximately) for $\ge$ 99.9
\% of all $\beta^{-}$ decaying atoms. The goal of this and following studies
is to correct such a situation at least for some light atoms. In general,
the results of experiments, in which the final state probabilities for
$\beta^{\pm}$ decaying atoms and molecules are measured, can be considered
as a very serious quantative test for modern theory of electron density
distribution in atoms and molecules.

Formally, the current theory of $\beta^{-}$ decay in atoms (and molecules)
is self-consistent and it does not include any unsolved problem. All
troubles of the current theoretical evaluations are mainly related with the
relatively low accuracy of the wave functions used in calculations. For
instance, in \cite{FrTa} we have calculated a large number of probabilities
for the `ground-state to ground-state' transitions. In fact, such
probabilities are now known for all atoms from He up to Ar \cite{FrTa}.
However, analogous calculations of the `ground-state to excited-states'
probabilities are significantly more difficult to perform, since for many
atoms/ions we do not have sufficiently accurate wave functions of the
excited states. Finally, the computed values of final state probabilities
for the excited states are not accurate. Furthermore, these values are
often oscillate, if the number of basis functions increases.

Analogous computations for the $\beta^{+}$ decays in atoms are even more
complicated. In particular, it is very hard to determine the final state
probabilities accurately, if a negatively charged ion is formed in the
result of the atomic $\beta^{+}$ decay. In such cases one needs to use
highly accurate methods which are specifically designed for accurate
computations of the negatively charged ions. In this study we have developed
such a method, and this allows us to determine the final state probabilities
in those cases when negatively charged ions are formed after the nuclear
$\beta^{+}$ decays in some few-electron atoms and ions. The probabilities
to form bound negatively charged ions which are computed below have never
been determined in earlier studies. Another interesting problem which has
never been discussed is the emission of the fast secondary electrons during
nuclear $\beta^{\pm}$ decays in many-electron atoms and molecules.

The present work has the following structure. In the next Section we discuss
a few numerical methods which are used to determine the bound state wave
functions of the incident and final states in few-electron atoms and ions.
Section III contains a brief discussion of the final state probabilities
computed for some $\beta^{-}$ decaying light atoms. Here we consider the He,
Li and Be atoms. Our present analysis is extensive and it includes a few
excited states in each of the final ion. In Section IV we determine the
`ground state to ground state' and `excited state to ground state' transition
probabilities for the $\beta^{+}$ decay in some light atoms. The final
atomic system in this case is a negatively charged ion. Emission of the
fast, secondary electron (or $\delta-$electrons) during the nuclear
$\beta^{\pm}$ decay in atoms are considered in Section V. The concluding
remarks can be found in the last Section.

\section{Method}

Let us assume that we have an $N-$electron atom which is described by its
bound state wave function $\Psi_i$, i.e. $H_{0} \Psi_i = E_i \Psi_i$, where
$H_{0}$ is the atomic Hamiltonian (see, e.g., \cite{LLQ}), $E_i$ is the
corresponding eigenvalue (or total energy, for short) and $\Psi_i$ is the
eigenfunction of the incident bound state which has a finite norm, i.e.
$\mid \Psi_i \mid^2$ = 1. Consider now a sudden change of the Hamiltonian of
atomic system. By sudden change we mean that the change in the original
Hamiltonian $H_{0}$ occurs in a time which is very short compared with the
periods of (atomic) transitions from the given state $i$ to other states.
The electron density distribution and the corresponding wave function cannot
change for such a short time and remain the same as before perturbation.
This means that after such a process we find the new atomic system with the
new Hamiltonian $H_f$, but with the old electron density distribution. Such
an electron density distribution is described by the old wave function
$\Psi_i$. The new Hamiltonian $H_f$ has a complete system of eigenfunctions,
i.e. $H_f \Phi^{(k)}_f = E_k \Phi^{(k)}_f$. Therefore, at the final stage of
the process we have only states with the wave functions $\Phi^{(k)}_f$. The
incident wave function $\Psi_i$ is now represented in the form of an
expansion $\Psi_i = \sum_{k} A_k \Phi^{(k)}_f$, where the coefficients $A_k$
can be considered as the transition (probability) amplitudes. The
corresponding probabilities $p_k = \mid A_k \mid^2$ determine the
probability to detect the final system in its state $\Phi^{(k)}_f$, if the
initial state of the system was described by the wave function $\Psi_i$.
Note that the system of notations used here correspond to the case of the
discrete spectra in both the incident and final atomic systems. In general,
the expansion $\Psi_i = \sum_{k} A_k \Phi^{(k)}_f$ must contain different
parts which represent the discrete and continuous spectra, respectively.

Thus, to determine the probability amplitudes $A_k$ we need to compute the
overlap integrals between two $N-$electron wave functions $\Psi_i$ and
$\Phi^{(k)}_f$ functions for different $k$, i.e.
\begin{equation}
 A_k = \int \Psi^{*}_i({\bf r}_1, \ldots, {\bf r}_N) \Phi^{(k)}_f({\bf r}_1,
 \ldots, {\bf r}_N) d^3{\bf r}_1 \cdot \ldots \cdot d^3{\bf r}_N \label{Int}
\end{equation}
In general, this value is complex, but the corresponding probabilities $p_k
= \mid A_k \mid^2$ are always real and their values are bounded between 0
and 1. As follows from Eq.(\ref{Int}) any of the final states must have the
same $L$ and $S$ quantum numbers as the incident state. Here and everywhere
below the notation $L$ designates the angular (electron) momentum of the
atom, while the notation $S$ denotes the total electron spin. Note that the
$L$ and $S$ quantum numbers are used in the non-relativistic
$LS$-classification scheme which is appropriate for light atoms and ions.
Briefly, we can say that the angular (electron) momentum of the atom and its
total electron spin are conserved during the nuclear $\beta^{-}$ decay. This
means that the original problem of determining the final state probabilities
in the case of $\beta^{-}$ decay in atoms is reduced to the construction of
highly accurate wave functions for the incident and final states with the
same $L$ and $S$ quantum numbers. This means the conservation of the angular
momentum ${\bf L}$ and total electron spin ${\bf S}$ during the nuclear
$\beta^{-}$ decay in many-electron atoms. In addition to these two quantum
numbers the spatial parity of the incident wave function is also conserved.

The conservation of the angular (electron) momentum $L$ and total electron
spin $S$ of the atom during the nuclear $\beta^{-}$ decay follows directly
from the perturbation theory. In fact, these conservation rules are not
fundamental, i.e. they are obeyed only in the lowest order approximations
upon $\alpha = \frac{e^2}{\hbar c} \approx \frac{1}{137}$, where $\alpha$ is
the fine structure constant. It can be shown that in higher order
approximations upon $\alpha$ the $L$ and $S$ quantum numbers do not conserve
(see discussion in Section V below). The leading correction to the
non-relativistic results (i.e. to the final state probabilities) is $\approx
\alpha^2 (\alpha Q)^2$, where $Q$ is the electric nuclear charge (in atomic
units). In light atoms such a correction is very small $\approx \alpha^4$
and can be ignored. In heavy atoms with $Q \approx 100$ the overall
contribution of this correction is substantially larger, but these atoms are
not considered in this work.

\subsection{Variational wave functions}

Numerical evaluations of the overlap integral, Eq.(\ref{Int}), require the
knowledge of highly accurate wave functions of the incident and final atomic
systems. To determine such wave functions for the ground and excited states
of different atoms and ions in this work we perform extensive calculations
of few-electron atomic systems. Then, by using our accurate wave functions
we determine the corresponding transition amplitudes and the final state
probabilities. This is the second step of our procedure. In this Section we
discuss the methods used to construct highly accurate wave functions of
few-electron atoms and ions. In general, the wave functions of the excited
states which have the same symmetry as the ground state can be found as the
solutions of the corresponding eigenvalue problem.

The energies of the different bound states are calculated by optimizing the
orbital exponents of the corresponding root of the eigenvalue equation.
Furthermore, our wave functions are simultaneously the eigenfunctions
of the angular momentum $\hat L^2$ and spin $\hat S^2$ operators,
respectively. Therefore, these eigenfunctions can be used in numerical
calculations of various bound state properties. The Slater orbitals are the
natural basis for all atomic calculations. In this study we also use the
basis of radial functions constructed from Slater orbitals.

To perform numerical computations of few-electron atoms and ions in this
study we apply the Hylleraas-Configuration Interaction method (Hy-CI) and
the Configuration Interaction method (CI) with Slater orbitals. Both these
methods are included in our package of computer codes. The Hy-CI method,
introduced by Sims and Hagstrom \cite{SimsJCP,Sims-Be}, combines the use of
orbitals with higher angular momentum (as in regular CI procedure) and
inclusion of the interelectronic distance $r_{ij}$ into the wave function
(as for Hylleraas-type trial wave functions). The Hy-CI and CI wave
functions for an $n$-electron systems are defined as:
\begin{equation}
\Psi =\sum_{k=1}^NC_k\Phi _k,\qquad \Phi _k=\hat{O}(\hat{L}^2)\hat{\mathcal{A%
}}\phi _k\chi   \label{wave}
\end{equation}
where $\Phi _k$ are symmetry adapted configurations, $N$ is the number of
configurations and the constants $C_k$ are determined variationally. The
operator $\hat{O}(\hat{L}^2)$ projects over the proper spatial space, so
that every configuration is eigenfunction of the square of the angular
momentum operator $\hat{L}^2$. $\hat{\mathcal{A}}$ is the $n$-particle
antisymmetrization operator, and $\chi $ is the spin eigenfunction:
\begin{equation}
\chi =\left[ (\alpha \beta -\beta \alpha )...(\alpha \beta -\beta \alpha
)\alpha \right] \label{spin}
\end{equation}
where for even electron systems the last $\alpha$ spin function is omitted.
The spatial part of the basis functions are Hartree products of Slater
orbitals:
\begin{equation}
\phi _k=r_{ij}^{\nu}\prod_{i=1}^n\phi _i(r_i,\theta _i,\varphi _i)
\label{Hartree}
\end{equation}
where the power $\nu$ takes the values $0$, or $1$. For $\nu=0$ the wave
function reduces effectively to a CI wave function. The basis functions
$\phi _k$, are the products of Slater orbitals defined as follows
\begin{equation}
\phi (\mathbf{r}) =r^{n-1}e^{-\alpha r} Y_l^m(\theta ,\phi ) \label{Slater}
\end{equation}
where $Y_l^m(\theta ,\phi )$ are the spherical harmonics. The phases used in
our definition of $Y_l^m(\theta ,\phi )$ correspond to the choice made by
Condon and Shortley \cite{Condon}, i.e.
\begin{equation}
Y_l^m(\theta ,\phi )=(-1)^m\left[ \frac{2l+1}{4\pi }\frac{(l-m)!}{(l+m)!}%
\right] ^{1/2}P_l^m(\cos {\theta })e^{im\phi }  \label{spherical}
\end{equation}
where $P_l^m(\cos {\theta })$ are the associated Legendre functions.

The integrals occurring in our calculation are up to four-electron integrals
in the Hy-CI method and two-electron integrals in the CI method. Expressions
for all these integrals are given in Refs. \cite{Ruiz3e,Ruiz4e,Sims3e}. The
calculation of the overlap between the wave functions of bound states
require only the usual two- and three-electron integrals.

Currently, the non-relativistic total energy of the ground state of helium
atom is known to very high accuracy (up to 40 decimal digits)
\cite{Nakatsuji-He}. Many excited $S$-, $P$-, $D$-, $F$-, etc, states in
two-electron helium atom have also been computed to high numerical accuracy
(see, e.g., \cite{Nakatsuji-exc,Drake,Sims-exc}). The ground $1^1S$-state of
the helium-like Li$^{+}$ ion (or ${}^{\infty}$Li$^{+}$ ion) has been
determined to high accuracy \cite{Nakatsuji-He2,Frolov-Li+}, while the $2^1S,
\ldots, 7^1S$ states in the Li$^{+}$ ion are known to significantly less
accuracy \cite{Perkins,Weiss,Pekeris}. Highly accurate calculations of the
excited $S$-states in the Li$^{+}$ ion higher than $7^1$S have never been
performed.

As a reference calculation in the case of helium-like two-electron atoms we
start with a Hy-CI energy of the ${}^{\infty}$He atom $-2.90372437699$ a.u.
This energy was obtained with the use of 820 configurations and a basis set
which included the $s$-, $p$-, $d$- and $f$-Slater orbitals $[18s,16p,16d,16f]$.
This total energy has uncertainty which is less than $1 \cdot 10^{-9}$
$a.u.$ The `optimal' exponent $\alpha$ = 2.9814 has been obtained by
optimizing 404 configurations constructed with a smaller basis
$[11s,11p,11d,11f]$. The best Hy-CI energy obtained with a single exponent
for the ground state of the He atom with the infinitely heavy nucleus is
$-2.90372437701$ a.u. (974 configurations). All these calculations have been
performed with the use of quadruple precision, or 30 decimal digits per
computer word. Some special measures have been taken to avoid any linear
dependence for this basis set.

The total energies of different bound $n^1S-$states in the Li$^{+}$ ion are
shown in Table I. In calculations of the overlap, which involve the wave
functions of the He atom and Li$^{+}$ ion, we have used the wave functions
for atom and ion with the same number of terms. The orbital exponents of
different states were always different. In fact, the orbital exponents of
every excited state have been optimized at several stages and used for the
larger basis (for more detail, see Table II). The optimal values of
exponents are shown in Table II. Every time when a new exponent has been
introduced in a series of calculations, a complete re-optimization has been
made. Currently, our best calculations have been performed with $820$
configurations, but the optimal exponents have been obtained in calculations
with a smaller basis $[14s,14p,14d,14f]$ ($680$ Hy-CI configurations).
The use of a single exponent (considering double occupancy of the orbitals)
for all configurations has been sufficient to obtain highly accurate
energies. The total energies obtained in this study for the $2^1S$-, $3^1S-$
and $4^1S$-states of the Li$^{+}$ ion are the lowest values obtained
to-date.

Note that our resulting wave functions derived after optimization are not
orthogonalized. Therefore the overlaps between configurations must be
determined. In turn, this problem is reduced to numerical calculation of the
overlap integrals. The symmetry adapted configurations have been constructed
for $S$-symmetry as $s(1)s(2)$, $s(1)s(2)r_{12}$, $p(1)p(2)$, $p(1)p(2)r_{12}$,
$d(1)d(2)r_{12}$, $f(1)f(2)$ and $f(1)f(2)r_{12}$. Using the short notation,
e.g., $p_0(1)p_0(2)=p_0p_0$, $p_{1}(1)p_{-1}(2)=p_{1}p_{-1}$, etc, we can
write the symmetry adapted configurations $pp$, $dd$ and $ff$ in the form:
\begin{eqnarray}
pp &=&p_0p_0-p_1p_{-1}-p_{-1}p_1  \nonumber \\
dd &=&d_0d_0-d_1d_{-1}-d_{-1}d_1+d_2d_{-2}+d_{-2}d_2  \nonumber \\
ff &=&f_0f_0-f_1f_{-1}-f_{-1}f_1+f_2f_{-2}+f_{-2}f_2-f_{-3}f_3-f_{-3}f_3
\label{Eqq8}
\end{eqnarray}

In Table II we also show the convergence of the energy with respect to
several truncated wave function expansions. The exponents used in every
calculation are given explicitly for each state. It was observed that for
the determination of higher excited states the diffuse functions are needed
and the wave functions expansions become larger. The total energies of the
first four excited states can be determined to the accuracy which is better
than $\pm 1 \cdot 10^{-6}$ $a.u.$ However, such an accuracy rapidly
decreases for the highly excited states. The value of the calculated overlap
integral, which includes the excited states of Li$^{+}$ and the ground state
of the He atom, substantially depends upon the overall accuracy of the
calculated energy. For low-lying states we have determined the overlaps with
overall accuracy $\approx$ 4-5 stable decimal digits. For higher states such
an accuracy decreases, but the absolute values of overlaps become very small
and tend to zero.

In calculations of the Li and Be atoms and corresponding isoelectronic ions
Be$^+$ and Li$^{-}$ we have used the wave functions constructed with the use
of L-S Configuration Interaction method. In the CI calculations we have used
double precision, which was been sufficient for our purposes. It was checked
by performing analogous calculations with the quadruple precision.
Calculations with double precision are significantly faster. The method used
for calculations and optimization of the orbital exponents is very similar
to the method used above for two-electron systems.

For the three-electron systems Li and Be$^+$ we use the Full Configuration
Interaction method (FCI). In such cases, therefore, there are no
configurations which have been either selected, or eliminated. We have used
a set of $s$-, $p$- and $d$-Slater orbitals and two exponents, considering
double occupancy of the orbitals. The exponents are the same for all
configurations. We have optimized the exponents for the smaller basis used,
i.e. n=3 [3s,2p,1d] or n=4 and employed them in the calculations with the larger
basis sets $n=5,6$. Eventually, the exponents in larger calculation are also
optimized. The configurations are symmetry adapted, and constructed
combining the two-electron configurations of Eq.(\ref{Eqq8}) with one
$s$-orbital. These configurations are $sss$, $spp$, $pps$, $sdd$ and $dds$.
The obtained energies have been determined with $\approx 1 \cdot 10^{-3} a.u.$
accuracy for the ground and excited states of the Li atom and Be$^+$ ion.
They are shown in Table III. The overlaps between the wave functions of the
ground state of Li atom and the ground and excited states of Be$^+$ have
been calculated numerically (see Table III). The values converge adequately
and the overlaps rapidly decrease for higher excited state.

For four-electron atomic systems we optimize the orbital exponents using a
small basis $n=4$ (this means $[4s3p2d1f])$, and use those exponents in
larger calculations with $n=5,6$. The configurations are grouped in blocks
for a given $n$ and according to the type (i.e. $ssss$, $sspp$, $ppss$,
$spps$, $\ldots$). Then the blocks of configurations have been filtered with
a threshold of average single configuration contribution of $\approx 1 \cdot
10^{-4}$. All blocks of configurations with small contribution to the total
energy have been eliminated after being tested. This could not produce any
substantial lost in the total energy. In reality, the corresponding error
was $\le 1 \cdot 10^{-3}$ $a.u.$ In addition, all configurations in our
calculations have been ordered according to their orbitals: $s$-, $p$-,
$d$-, and $f$-orbitals, and within these groups by approximately energetic
order.

As the ground state of the Be atom is also a ${}^1S$-state, the
configurations can be constructed combining the two-electron symmetry
adapted configurations of Eq.(\ref{Eqq8}). Resulting configurations are:
$ssss$, $sspp$, $spps$, $ppss$, $pppp$, $ssdd$, $sdds$, $ddss$, $sppd$,
$dpps$, $sdpp$, $ppdd$, $pddp$, $ddpp$, $ssff$, $ddff$, $ffff$. A set of two
exponents (double occupancy of the shells) has been used. With this
restriction, the configurations showed above represent all possible cases
that can be formed. Nevertheless, the configurations $
ddff,ffff$ have been eliminated because their contributions were less than
the threshold. An additional configuration type of S-symmetry $sppd$ and its
permutations $dpps$ and $sdpp$ contribute considerably to the energy
calculations on four-electron systems, but not in three-electron ones, where
they contribute $\le 1 \cdot 10^{-4}$ $a.u.$. This configuration is somehow
more complex:
\begin{eqnarray}
sppd &=& sp_0p_0d_0 + sp_1p_1d_{-2} + sp_{-1}p_{-1}d_2 +  sp_1p_{-1}d_0 \nonumber \\
     &+& sp_{-1}p_1d_0 - sp_1p_0d_{-1} - sp_{-1}p_0d_1 - sp_0p_1d_{-1} - sp_0p_{-1}d_1
\label{Eqq9}
\end{eqnarray}

Table IV contains the probability amplitude and final state probability for
the $\beta^{+}$-decay of the four-electron Be atom into four-electron
Li$^{-}$ ion. In this case in numerical calculation of the overlap we
follow the same method of calculation used above for two-electron systems.
However, for four-electron atomic systems no Hy-CI terms have been included.
We are planning to include such terms in future studies. Since the computed
CI energies are known to the accuracy $\pm 1 \cdot 10^{-3}$ $a.u.$, then
we restrict here our calculations to the lowest three $S$-states in the
incident Be atom. The calculated ground state energy of the Li$^{-}$ ion is
$-7.498913845101$ a.u. (for 2155 CI configurations), and it is close
to the best-to-date value $-7.50058250$ a.u. \cite{Frolov-Li-} known in the
literature for this system. The calculated total energy of the ground state
of Be atom is $-14.665206189$ a.u. is very close to the best results of
recent calculations $-14.667356486$ a.u. \cite{Adam-Be}. This value agrees
very well with the value $-14.66544500$ a.u. calculated by Bunge with
approximately the same basis \cite{Bunge-Be}. As expected the excited states
of the Be atom can be determined with less accuracy than the ground state.
The calculated total energies together with the overlaps between wave
functions of the ground/excited states of the Be atom and the ground state
of the Li$^{-}$ ion can be found in Table V.

Finally, the `ground state to ground state' transition probability for
$\beta^{-}$-decaying Be atom (to B$^+$ ion) can be found in Table VI. The
reference ground state energies for the Be atom are given in Table V.
Note that our ground state energy of the B$^+$ ion has an overall accuracy
which is better than $\pm 1 \cdot 10^{-3}$ $a.u.$

\section{Results for $\beta^{-}$ decaying light atoms}

As we mentioned above in this study we consider the $\beta^{-}$ decays in a
number of few-electron atoms He, Li, and Be. In all our calculations we
assume that before the nuclear $\beta^{-}$ decay each of the atoms was in
its ground state (except calculations shown in Table V). Furthermore, the
probability of direct electron ionization during $\beta^{-}$ decay was
assumed to be small. Its contribution is essentially ignored in this study.
Numerical evaluation of the corresponding small correction can be found in
Section V below. Briefly, this means that all ions which are formed after
the nuclear $\beta^{-}$ decay contain the same number of electrons as the
original atoms. In other words, all final state probabilities can be
determined with the use of Eq.(\ref{Int}) where the overlap integral
contains two $N-$electron wave functions. For instance, the nuclear
$\beta^{-}$ decay of the He atom produces the two-electron Li$^{+}$ ion. If
the incident He atom was in its ground $1^1S(L = 0)-$state, then, in respect
with the conservation rules formulated above, the final two-electron
Li$^{+}$ ion will be in one of its bound $n^1S(L = 0)-$states, where $n = 1,
2, 3, \ldots$, or in an unbound state. In this study we consider the bound
$n^1S(L = 0)-$states in the Li$^{+}$ ion up to $n = 8$. The transition
amplitudes $A_{g \rightarrow n}$ and corresponding probabilities $p_{g
\rightarrow n} = \mid A_{g \rightarrow n} \mid^2$ for the nuclear
$\beta^{-}$ decay of the He atom can be found in Table I. Table I also
contains the total energies of all $n^1S(L = 0)-$states (for $n = 1, 2,
\ldots, 8$) in the Li$^{+}$ ion. These energies indicate, in principle, the
overall quality of the bound state wave functions used in our calculations
of the overlap integrals, Eq.(\ref{Int}). The wave function of the ground
$1^1S(L = 0)-$state in the incident He atom corresponds to the energy $E$ =
-2.90372437701 a.u. which is very good for Hy-CI wave function with $N \le
974$ terms.

Note that there are a few simple rules which must be obeyed, in principle,
for any distribution of the final state probabilities $p_{g \rightarrow n}$
obtained in numerical calculations. For simplicity, let us restrict
ourselves to the cases when all final states are also bound and each of
these states is labeled with the integer quantum number $n (n \ge 0)$. This
quantum number $n$ is often called the `excitation number' and/or `index of
excitation'. The value $n = 0$ corresponds to the ground state in
few-electron atom, i.e. $n = g$. The first rule for probability distribution
is simple and states that the numerical values of such probabilities rapidly
decrease, if the excitation number $n$ increases, i.e. it must be $p_{g
\rightarrow n} > p_{g \rightarrow (n+1)}$ for an arbitrary $n$ ($n \ge 0$).
In reality this inequality is even stronger, i.e. $p_{g \rightarrow (n+1)}
\ll p_{g \rightarrow n}$. In some actual calculations one can find an
opposite inequality for the final state probabilities. Usually, it is
directly related with very slow convergence rate(s) for the wave functions
of the incident and final atomic systems. Numerical values of these final
state probabilities cannot be used in actual applications. They must be
improved in future calculations with better convergent basis sets. The only
expectation from this rule can be found in those cases, when the ground state
wave function of the incident system and the trial wave function of one of
the excited states of the final ion are almost orthogonal to each other. The
final state probability is a very small value for such an excited state.
In many cases, it is directly follows from an additional symmetry of the
basis functions used to construct the variational wave functions.

The second rule states that the sum of all partial probabilities must
converge to the value which exceeds $\approx$ 0.75 (if the initial system
was a neutral atom), but always less than unity. In fact, the difference
\begin{equation}
 P_{ion}(g) = 1 - \sum^{N_{max}}_{n=1} p_{g \rightarrow n} \label{sum}
\end{equation}
is the total probability of electron ionization (from the ground state $g$)
during the nuclear $\beta^{-}$ decay in a neutral atom. Ionization means
that after $\beta^{-}-$decay the total number of bound electrons decreases
by unity. It is clear that the sum in Eq.(\ref{sum}) must be infinite, i.e.
$N_{max} = \infty$. In actual computations, however, there is a problem of
slow convergence for the wave functions of highly excited bound states.
This means that in actual cases the sum Eq.(\ref{sum}) is usually finite.
The actual maximal value of $N$, in $N_{max}$, in Eq.(\ref{sum}) is
determined by the first rule mentioned above, i.e. in the sum,
Eq.(\ref{sum}), we can use only those bound states for which the inequality
$p_{g \rightarrow n} > p_{g \rightarrow (n+1)}$ is obeyed. The approximate
value of $P_{ion}$ determined for the nuclear $\beta^{-}$ decay in the He
atom with the use of our results from Table I is $P_{ion} \approx 0.108$. In
other words, the one-electron Li$^{2+}$ ions are formed in $\approx$ 10.8 \%
of all $\beta^{-}$ decays of the He atoms. In actual experimental conditions
these ions can be observed in the $\beta^{-}$ decays of the ${}^6$He atoms.
The half-life of the ${}^6$He atom against such a $\beta^{-}$ decay is
$\approx$ 0.82 $sec$.

In general, the method described above can be used to determine the total
probability of ionization during the nuclear $\beta^{-}$ decay in any
neutral atom. It is very simple and has many advantages in comparison with
the so-called `direct' methods. In these direct methods the wave functions
of the out-going electron and double-charged final ion must be explicitly
constructed. Then one needs to compute the overlap integral between the
product of these two wave functions and wave function of the incident atom.
This step corresponds to the sudden approximation used above. However, in
actual calculations we cannot assume that the out-going electron is always
in the $s-$wave. Briefly, this means that we need to include many
configurations in which the final (free) electron moves in the $p-, d-, f-,
\ldots$ waves, while the double-charged final ion is in one of its $P-, D-,
F-, \ldots$ states, respectively. If the incident atom was in one of its
$S-$states, then only the $sS-, pP-, dD-$, etc, configurations for the final
system must be used in calculations. The total energies some of these
configurations are close to each other. To reach a `realistic' accuracy one
needs to consider a very large number (up to few dozens) of different
configurations (with different $L$) in the final system. In general, each of
these computations is not easy to conduct with relatively high accuracy.
This significantly complicates all direct calculations of the ionization
probabilities.

An interesting and actual question is the convergence of computational
results obtained for the transition amplitudes and transition probabilities.
Recently, a number of papers have been published about nuclear $\beta^{-}$
decay in different atoms and ions. In all these works it was assumed that
the determined transition amplitudes and corresponding probabilities are
`exact', i.e., they will not change noticeably in similar future
calculations. In many cases, however, the following calculations show that
such results were not exact and overall changes in some cases are relatively
large. In particular, all calculations of the transition amplitudes and
transition probabilities performed with the use of Hartree, Hartree-Fock and
Hartree-Fock-CI methods cannot be considered as very accurate unless some
additional measures have been taken. In this study we decided to analyze
this problem in detail. The result of our analysis can be found in Table II
where various transition amplitudes and transition probabilities are
determined with the different number(s) of basis functions.

As follows from Table II our method provides a very good convergence rate
for the ground and low-excited $n^1S(L = 0)-$states in the Li$^{+}$ ion. For
the excited $n^1S(L = 0)-$states with $n \ge 6$ the overall convergence rate
drops drastically. In such cases to keep the overall accuracy of our
calculations of the corresponding overlap integrals we need to use larger
numbers of basis functions. In general, it is very hard to
compute transition probabilities for highly excited (bound) states of the
final atomic system. On the other hand, the numerical values of these
probabilities decrease rapidly when the `excitation number' $n$ increases.
Therefore, by using a few known transition probabilities into the lowest
bound states of the final system we can accurately evaluate the total
`ground state to bound states' probability and total `ionization
probability' for an arbitrary $\beta^{-}$-decaying atom.

Our results obtained for the atomic transition amplitudes and corresponding
transition probabilities for the nuclear $\beta^{-}$ decay in the Li atom
can be found in Table III. In these calculations we assume that the original
Li atom was in its ground (doublet) $1^2S-$state. Due to the conservation of
the $L$ and $S$ quantum numbers the final Be$^{+}$ ion will be in one of its
bound (doublet) $n^{2}S-$states. The final states probability amplitudes and
corresponding probabilities have been computed with the use Eq.(\ref{Int}).
The `ground state to ground state' probability and the corresponding
transition amplitude for the $\beta^{-}$-decaying B atom are shown in Table
VI. Note that for all elements discussed in this study our computed
`ground-state to ground-state' probabilities coincide well with the
corresponding results from \cite{FrTa}. However, if the final ion is in one
of its excited states, then our current results have substantially better
accuracy. This is directly related with the better overall accuracy of our
current wave functions.

The knowledge of the final state probabilities allows one to predict the
excitations of the final atomic fragment, i.e. in the final atom/ion. In
general, any excited state in few-electron atom decays with the emission of
a few optical quanta. These transitions produce an unique spectrum of
post-decay optical radiation. By using the computed final state
probabilities we can estimate the spectrum and intensity of the post-decay
optical radiation which is observed for some time $\tau$ (usually $\tau
\approx 1 \cdot 10^{-9} - 1 \cdot 10^{-2}$ $sec$) after the nuclear
$\beta^{-}$ decay. In the case of $\beta^{-}$ decaying ${}^6$He atom (from
its ground state) the post-decay optical radiation corresponds to the
chain of optical transitions from the final $n^1S$-state of the Li$^{+}$ ion
into its ground $1^1S$-state. For instance, for the $3^1S$-state in the
Li$^{+}$ ion this chain of dipole transitions is: $3^1S \rightarrow 2^{1}P
\rightarrow 1^1S$. Various collisions between Li$^{+}$ ions and He/Li atoms
and possible electron capture by the Li$^{+}$ ion must also be taken into
account. The arising (optical) spectrum of post-decay radiation is very
complex, but it can be studied, in principle, with the use of theoretical
and current experimental methods.

\section{Formation of the negatively charged ions during the $\beta^{+}$
decay in few-electron atoms}

Formation of the negatively charged ions (or anions) during the nuclear
$\beta^{+}$ decay in many-electron atoms is a very interesting experimental
problem. On the other hand, it is very interesting to evaluate the
corresponding final state probabilities by using our computational methods
described above. It is clear $a$ $priori$ that such probabilities can be
found with the use of the sudden approximation (exactly as it was made above
for the nuclear $\beta^{-}$ decay). Formally, in the case of the nuclear
$\beta^{+}$ decay in many-electron atoms one needs to determine the same
overlap integral, Eq.(\ref{Int}), between the incident and final
$N$-electron wave functions. This is exactly the same procedure as described
above for the nuclear $\beta^{-}$ decay, but actual computations of the
overlap integrals, Eq.(\ref{Int}), is a significantly more complicated
problem in those cases when the negatively charged ions are involved. The
first complication follows from the experimental fact that many atoms do not
form stable negatively charged ions. However, if such negatively charged
ions are stable, then the construction of highly accurate variational wave
function(s) for these ions is a very hard problem. Briefly, this means that
the final state probabilities obtained for the nuclear $\beta^{+}$ decay in
many-electron atoms are not very reliable, if they have been determined for
the negatively charged ions.

Nevertheless, in this study we have determined probabilities for the
ground-state (atom) to ground-state (negative ion) transition for the
nuclear $\beta^{+}$ decay in some light atoms. Note that each of the
negatively charged atomic ions have either one bound (ground) state, or no
bound states et al. In particular, we consider a possibility to form the
${}^7$Li$^{-}$ ion during the nuclear $\beta^{+}$ decay of the ${}^7$Be
nucleus. It should be mentioned that more than 99 \% of all ${}^7$Be nuclei
decay by the electron capture. If the $K-$electron capture in the ${}^7$Be
atom occurs, then the Li$^{-}$ ion cannot be formed. However, any
experimental observation of the Li$^{-}$ ions from decaying ${}^7$Be nuclei
will be an actual indication of the competing $\beta^{+}$ decay. As follows
from Table IV the total probability to form the bound Li$^{-}$ ion during
such a decay is evaluated as $\approx$ 0.2065. This means that the Li$^{-}$
ions will form in $\approx$ 20.65 \% of all nuclear $\beta^{+}$ decays of Be
atoms, i.e. in one of five such decays we can observe the negatively charged
Li$^{-}$ ion. Another interesting result can be found in Table V. As follows
from that Table the probability to form the bound Li$^{-}$ ion is larger
$\approx$ 35.5 \% in those cases, when the incident Be atom was in its
excited $2^1S-$state. It indicates clearly that the distribution of the
final state probabilities in those cases when the negatively charged ions
are formed is very different from the known distributions of
$\beta^{-}$-decaying neutral atoms.

Another interesting $\beta^{+}$-decaying atomic system with small number of
electrons is the ${}^{11}$B$^{-}$ ion. This ion arises during the nuclear
$\beta^{+}$ decay of the ${}^{11}$C atom ($\tau_{\beta^{+}}({}^{11}$C)
$\approx 20.4$ min). Very likely, the formation of the ${}^{11}$B$^{-}$ ion
will be the first actual experiment which can confirm the direct formation
of the negatively charged ions during the nuclear $\beta^{+}$ decay.

The formation of negatively charged ions during the nuclear $\beta^{+}$
decay has a great theoretical interest, since the probability to form such
ions directly related to the change in distribution of the outer most
electron(s). Furthermore, the density distribution of the outer most
electron(s) for all negatively charged ions is very similar to each other.
As is well known (see, e.g., \cite{Ost}) the radial wave function $R(r)$ of
an arbitrary $N-$electron atomic system with the nuclear charge $Q$ at large
$r$ has the following asymptotic form
\begin{equation}
 R^{Q}(r) \sim r^{b-1} \cdot \exp(-t r) = r^{\frac{Q^{*}}{t}-1} \cdot
 \exp(-t r) \label{asym1}
\end{equation}
where $t = \sqrt{2 I}, b = Q^{*}/t$ and $Q^{*} = Q - N + 1$. Here the
notation $I$ stands for the first ionization potential. For negatively
charged ions $Q^{*} = Q - N + 1 = 0$ and $R^{Q}(r) = \frac{1}{r} \exp(-t
r)$, i.e., it does not depend explicitly upon $Q$. This substantially
simplifies all following evaluations and makes them universal for all
negatively charged ions. In particular, we can expect that the
total probabilities of negative ions formation will accurately be
represented by one relatively simple formula which contains only a few
parameters. This means that, if we know such probabilities for some of the
negatively charged ions, then we can accurately predict analogous values
for other similar ions.

\section{Emission of the fast $\delta-$electrons during the nuclear
$\beta^{-}$ decay in atoms}

The sudden approximation used above allows one to determine the final state
probabilities for the $\beta^{\pm}$ decays in many-electron atoms. Briefly,
the analysis of atomic excitations is reduced to the description of changes
in electron-density distribution produced by a sudden change of the nuclear
electric charge $Q \rightarrow Q \pm 1$. The electronic/positronic nature of
the $\beta^{\pm}$ decay is not critically important for our method. However,
the sudden approximation is true only in the lowest order approximations upon
the fine structure constant $\alpha$. This means that, if we are interested
in highly accurate results for the final state probabilities, then we need
to consider and evaluate the corresponding correction(s). The leading
contribution comes from the lowest-order correction on electron-electron
scattering which is $\approx \alpha^2 (\alpha Q)^2$. In heavy atoms with $Q
\approx 100$ such a correction is relatively large $\approx \alpha^2$, but
in light, few-electron atoms it is significantly smaller $\approx \alpha^4$.
Nevertheless, this
correction describes the new phenomenon, i.e. the emission of the fast
secondary electrons, which are traditionally called the $\delta-$electrons.
Let us discuss this phenomenon in detail. As is well known from Quantum
Electrodynamics (see, e.g., \cite{AB}, \cite{Grei}) the differential
cross-section of the electron-electron scattering is written in the form
\begin{eqnarray}
 d\sigma = 2 \pi \alpha^4 a^{2}_{0} \frac{dx}{\gamma^2 - 1} \Bigl[ 1 +
 \frac{(\gamma - 1)^2 \gamma^2}{x^2 (\gamma - 1 - x)^2} -
 \frac{2 \gamma^2 + 2 \gamma - 1}{x (\gamma - 1 - x)} \Bigr] \label{cross}
\end{eqnarray}
where $a_0$ is the Bohr radius, $\gamma$ is the $\gamma$-factor of the
$\beta$-electron emitted from the nucleus, while the parameter $x$ is the
energy lost by the $\beta$-electron (or gained by the atomic electron $a$),
i.e.
\begin{equation}
 x = \frac{\epsilon_{\beta} - \epsilon^{\prime}_{\beta}}{m_e c^2} =
 \frac{\epsilon^{\prime}_{a} - \epsilon_{a}}{m_e c^2} \label{eq14}
\end{equation}
where the superscript $\prime$ designates the particle after the process.
It is usually assumed that one of the two electrons (atomic electron in our
case) was at rest before electron-electron collision/scattering, i.e.
$\epsilon_{a} = m_e c^2$.

The formula Eq.(\ref{cross}) is the closed expression for the differential
cross-section of electron-electron scattering which depends upon the
parameter $x$, Eq.(\ref{eq14}), and $\gamma-$factor of the $\beta^{-}$
electron. As follows from Eq.(\ref{cross}) the probability to
observe/produce a fast $\delta$-electron during the nuclear
$\beta^{-}$-decay is very small in comparison with `regular' atomic
processes, since it contains an additional factor $\alpha^4 \approx 2.83
\cdot 10^{-8}$. Note also that the formula, Eq.(\ref{cross}), is derived for
a free electron which is located at the distance $a_0$ from atomic nucleus.
The actual $K-$electrons in heavy atoms are significantly closer to the
nucleus than electrons from outer electron shells. The effective radius of
the $K-$electron shell is smaller than $a_0$ in $\approx Q^2$ times. This
means that the factor $2 \pi \alpha^4 a^{2}_{0}$ in formula,
Eq.(\ref{cross}), must be multiplied by an additional factor $Q^2$. For
light atoms considered in our study the overall probability to observe the
emission of the fast $\delta-$electrons during the nuclear $\beta^{-}$
decay is very small. The situation changes for heavy atoms with $Q \approx$
90 - 100, but such atoms are not discussed here.

The emission of the fast $\delta-$electrons can also be observed during the
nuclear $\beta^{+}$ decay in many-electron atoms. In such a case, the
formula for the cross-section of electron-positron scattering takes the form
\cite{AB}, \cite{Grei}
\begin{eqnarray}
 d\sigma = 2 \pi \alpha^4 a^{2}_{0} \frac{dx}{\gamma^2 - 1} \Bigl[
 \frac{\gamma^2}{x^2} - \frac{2 \gamma^2 + 4 \gamma + 1}{(\gamma + 1) x}
 + \frac{3 \gamma^2 + 6 \gamma + 4}{(\gamma + 1)^2} -
 \frac{2 \gamma}{(\gamma + 1)^2} \Delta + \frac{1}{(\gamma + 1)^2}
 \Delta^2 \Bigr] \label{cross1}
\end{eqnarray}
where $\gamma$ is the $\gamma$-factor of the positron emitted from the
nucleus, while all other notations are the exactly same as in
Eq.(\ref{cross}). Note again that in light atomic systems the cross-section
Eq.(\ref{cross1}) is very small. In heavy atoms the situation changes and
in one of $\approx$ 17,000 nuclear $\beta^{+}$ decays we can also observe
the emission of the fast $\delta-$electron.

\section{Discussion and Conclusion}

We have considered atomic excitations arising during the nuclear $\beta^{-}$
decay. For some light few-electron atoms such final state
probabilities and the total ionization probabilities have been determined
numerically to a very good numerical accuracy. Our interest to light atoms
is directly related to the fact that currently the highly accurate wave
functions of the ground and 6 - 8 low-lying excited states can only be
constructed for some few-electron atoms and ions. Consideration of the 6 - 8
bound states in the final atomic system allows us to perform a complete
analysis of atomic excitations during the nuclear $\beta^{-}$ decay. We also
consider the formation of negatively charged ions during the nuclear
$\beta^{+}$ decay. By using our highly accurate wave functions for the
negatively charged ions we have determined the `ground state to ground
state' probabilities for some nuclear $\beta^{+}$ decays in which such
negatively charged ions are formed (or can be formed).

It should be mentioned that for the first time atomic excitations during the
nuclear $\beta^{\pm}$ decay were observed in 1912 (all earlier references on
this matter can be found, e.g., in \cite{Mig1}, \cite{Finb}, \cite{Skoro},
\cite{Schw}). In general, atomic and molecular excitations arising during
the nuclear $\beta^{\pm}$ decay have many interesting aspects for
theoretical study and experimental investigation. Analysis of the direct
atomic excitations in earlier studies was substantially restricted by the
use of non-accurate atomic wave functions. In this study we have applied
highly accurate wave functions for all few-electron atoms and ions. The
overall accuracy of our predictions for many excited states has increased
significantly. In future studies we want to extend our analysis to atomic
systems with more electrons. A separate goal will be a consideration of
different atomic (and molecular) excitations, analysis of the post-decay
radiation, etc.

Note that the final state probabilities determined above for a number of
$\beta^{-}$-decaying light atoms can also be used as important numerical
tests for other similar values needed in the analysis of various nuclear
reactions in few-electron atoms/ions. For instance, for exothermic nuclear
$(n;t)-, (n;p)-$ and $(n;\alpha)-$reactions in few-electron atoms/ions
\cite{FrWa} one needs to determine the numerical value of the following
integral
\begin{equation}
 A_k({\bf V}) = \int \Psi^{*}_i({\bf r}_1, \ldots, {\bf r}_N)
 \exp[\imath {\bf V} \cdot (\sum^{N}_{i=1} {\bf r}_i)] \Phi^{(k)}_f({\bf
 r}_1, \ldots, {\bf r}_N) d^3{\bf r}_1 \ldots d^3{\bf r}_N
 \label{Int1}
\end{equation}
where $N$ is the total number of bound electrons (here we assume that $N$
does not change during the nuclear reaction), while ${\bf V}$ is the nuclear
velocity in the final state, i.e. after the nuclear reaction. Note that in
the limit ${\bf V} \rightarrow 0$ the value $A_k({\bf V})$ from
Eq.(\ref{Int1}) converges to the $A_k$ value from Eq.(\ref{Int}). This
explains why the final state probabilities determined by Eq.(\ref{Int}) are
often considered as the `nucleus-at-rest' limit of atomic probabilities
obtained for more general nuclear reactions.

In conclusion, we want to note that this work opens a new avenue in the
analysis of atomic excitations during the nuclear $\beta^{\pm}-$decay in
atoms and molecules. Currently, many aspects of this problem are of
significant experimental and theoretical interest. In particular, the study
of atomic excitations arising in the nuclear $\beta^{\pm}$ decay can improve
our understanding of many atomic and QED processes. Furthermore, the
complete and accurate analysis of atomic excitations during various nuclear
reactions and processes is a complex problem which requires an extensive
development of new numerical methods and algorithms. It should be mentioned
that a sudden change of the electric charge of atomic nucleus and following
changes in the electron density distribution during the nuclear $\beta^{-}$
decay must be a great interest for the density functional theory (DFT) of
atoms and molecules. Note also that analysis of possible molecular
excitations arising during the nuclear $\beta^{-}$ decay in molecules is a
significantly more complicated problem, than analogous problem for atoms.
Nevertheless, some useful conclusions about different excitations in
molecular systems can be made and corresponding probabilities can be
evaluated numerically. In fact, in the last five-seven years we have
achieved a remarkable progress in understanding of atomic excitations during
various nuclear processes, reactions and decays. Unfortunately, except a
very few experimental papers published as a rule years ago (see, e.g.,
\cite{Carl1}, \cite{Carl2} and \cite{Scie}) the current theory of atomic
excitations during various nuclear reactions has no experimental support.
This is a very strange situation, since all required (atomic) experiments
are very easy to perform. Currently, we can only hope that this our work
will stimulate some experimental activity in the area.

\newpage

\begin{center}
    {\bf Acknowledgements}
\end{center}

It is a pleasure to thank James S. Sims (NIST) for fruitful discussions on
the Hy-CI method. One of us (AMF) wants to thank the University of Western
Ontario and the Dean of Science of UWO Prof. David M. Wardlaw for financial
support. MBR would like to thank Carlos Bunge for advises on the CI method,
and Prof. Peter Otto of the University Erlangen-N\"urnberg for supporting
this project.

\newpage


{\footnotesize
\begin{table}[tbp]
\caption{Table have been performed with the use of Hy-CI wave functions with
$s$-, $p$-, $d$-, and $f$-Slater orbitals. The He atom wave function
($1^1S$-state) used in these calculations corresponds to the energy
$-2.903 724 376 99(10)^a$ a.u. with 820 configurations. }
\begin{center}
{\footnotesize
\begin{tabular}{crccclc}
\hline\hline
State of Li$^+$ &  N & $\qquad$ Amplitude $\qquad$ & $\qquad$ Probability $\qquad$ & $\qquad$ Energy Li$^+$ (a.u.) &
Ref. Ener. (a.u.) & Ref. \\ \hline
 $1^1$S  & 820 & $\;$ 0.8417 9425 437 & $\;$ 0.7086 1757 & -7.2799 1340 7455 & -7.2799 1341 267$^b$ & \cite{Frolov-Li+} \\
 $2^1$S  & 820 & $\;$ 0.3865 0278 074 & $\;$ 0.1493 8440 & -5.0408 7674 3824 & -5.0408 7673 101 & \cite{Pekeris} \\
 $3^1$S  & 820 & $\;$ 0.1363 0370 050 & $\;$ 0.0185 7870 & -4.7337 5581 3588 & -4.7337 32 & \cite{Weiss} \\
 $4^1$S  & 820 & $\;$ 0.0790 1458 630 & $\;$ 0.0062 4330 & -4.6297 8349 2764 & -4.6297 78 & \cite{Perkins} \\
 $5^1$S  & 820 & $\;$ 0.0536 8725 837 & $\;$ 0.0028 8232 & -4.5824 2193 2563 & -4.5824 24 & \cite{Perkins} \\
 $6^1$S  & 820 & $\;$ 0.0401 7188 870 & $\;$ 0.0016 1378 & -4.5568 7765 0705 & -4.5569 51 & \cite{Perkins} \\
 $7^1$S  & 820 & $\;$ 0.0328 7934 851 & $\;$ 0.0010 8105 & -4.5408 7695 5351 & -4.5416 92 & \cite{Perkins} \\
 $8^1$S  & 820 & $\;$ 0.0432 0219 907 & $\;$ 0.0018 6643 & -4.5285 0740 1021 &  &  \\
\hline\hline
\end{tabular}
}
\end{center}
\par
{\footnotesize
\footnotetext[1]{{\footnotesize Reference energies: non-relativistic total
energy of the helium atom (ground state):
-2.9037 2437 7034 1195 8311 034$^e$ a.u. (variational) and
-2.9037 2437 7034 1195 8311 20(7)$^e$ a.u. (asymptotic) \cite{Frolov-He}%
. The best-to-date non-relativistic energy for the He atom contains over 40
stable decimal digits -2.9037 2437 7034 1195 8311 1592 4519 4404 4466 9669 a.u.
\cite{Nakatsuji-He}. }} }
\par
{\footnotesize
\footnotetext[2]{{\footnotesize The computed energy is -7.2799 1341 2669 30596
a.u. \cite{Frolov-Li+}. The lowest computed energy is
-7.2799 1341 2669 3059 6491 9459 2210 0661 1682 a.u. \cite
{Nakatsuji-He}.}}  }
\end{table}

\newpage


\begin{table}[tbp]
\caption{Convergences for the transition amplitudes, final state
probabilities and total energies for different $n^1S-$states of the
Li$^{+}$ ion.}
\begin{center}
\scalebox{0.72}{%
\begin{tabular}{ccccccc}
\hline\hline
State of Li$^+$ & N & $\qquad$ exp.$\qquad$ &  $\qquad$ Energy (a.u.)  &  Ref. Ener. (a.u.) & $\qquad$ Amplitude $\qquad$ & $ \qquad$ Probability \\
\hline
1 $^1$S & 365 & 2.520933 & -7.2799 1338 7919 &               & 0.8417 8267 952 & 0.7085 9808 \\
1 $^1$S & 631 & 2.520933 & -7.2799 1340 7024 &               & 0.8417 8267 547 & 0.7085 9807 \\
1 $^1$S & 820 & 2.520933 & -7.2799 1340 7455 & -7.2799134127 & 0.8417 9425 437 & 0.7086 1757 \\
\hline
2 $^1$S & 365 & 2.310125 & -5.0408 7670 8765 &               & 0.3864 9574 413 & 0.1493 7896 \\
2 $^1$S & 631 & 2.310125 & -5.0408 7674 3717 &               & 0.3864 9569 073 & 0.1493 7892 \\
2 $^1$S & 820 & 2.310125 & -5.0408 7674 3824 & -5.0408767    & 0.3865 0278 074 & 0.1493 8440 \\
\hline
3 $^1$S & 365 & 1.914217 & -4.7336 2880 1162 &               & 0.1367 1368 082 & 0.0186 9063 \\
3 $^1$S & 433 & 1.914217 & -4.7337 1118 0038 &               & 0.1364 6254 194 & 0.0186 2203 \\
3 $^1$S & 631 & 1.914217 & -4.7337 5112 1179 &               & 0.1363 2291 461 & 0.0185 8394 \\
3 $^1$S & 820 & 1.703409 & -4.7337 5581 3588 & -4.733732     & 0.1363 0370 050 & 0.0185 7870 \\
\hline
4 $^1$S & 365 & 1.821455 & -4.6183 8011 5848 &               & 0.0966 2820 138 & 0.0093 3701 \\
4 $^1$S & 433 & 1.728693 & -4.6256 0911 1417 &               & 0.0874 8471 613 & 0.0076 5358 \\
4 $^1$S & 463 & 1.728693 & -4.6275 0363 0034 &               & 0.0842 7564 245 & 0.0071 0238 \\
4 $^1$S & 631 & 1.728693 & -4.6285 8858 0287 &               & 0.0820 9716 663 & 0.0067 3994 \\
4 $^1$S & 722 & 1.307501 & -4.6297 7002 4419 &               & 0.0789 7160 683 & 0.0062 3651 \\
4 $^1$S & 820 & 0.978434 & -4.6297 8349 2764 & -4.629778     & 0.0790 1458 630 & 0.0062 4330 \\
\hline
5 $^1$S & 463 & 1.544017 & -4.5640 1993 0704 &               & 0.0781 3224 403 & 0.0061 0465 \\
5 $^1$S & 548 & 1.544017 & -4.5641 3928 7063 &               & 0.0780 3507 816 & 0.0060 8947 \\
5 $^1$S & 631 & 1.544017 & -4.5702 4950 5974 &               & 0.0726 2358 467 & 0.0052 7418 \\
5 $^1$S & 722 & 1.307501 & -4.5795 6274 1508 &               & 0.0607 3396 667 & 0.0036 8861 \\
5 $^1$S & 792 & 1.307501 & -4.5815 4166 3792 &               & 0.0599 1421 622 & 0.0035 8971 \\
5 $^1$S & 820 & 0.891026 & -4.5824 2193 2563 & -4.582424     & 0.0536 8725 837 & 0.0028 8232 \\
\hline
6 $^1$S & 548 & 1.610647 & -4.4548 0210 9793 &               & 0.0938 8184 073 & 0.0088 1380 \\
6 $^1$S & 631 & 1.610647 & -4.4780 5225 5218 &               & 0.0880 5122 442 & 0.0077 5302 \\
6 $^1$S & 722 & 1.399839 & -4.5164 8983 4858 &               & 0.0756 3927 569 & 0.0057 2130 \\
6 $^1$S & 792 & 1.399839 & -4.5424 8852 3968 &               & 0.0572 2105 469 & 0.0032 7425 \\
6 $^1$S & 820 & 0.891026 & -4.5557 8489 1486 &               & 0.0442 2008 877 & 0.0019 5542 \\
6 $^1$S & 820 & 0.691701 & -4.5568 7765 0705 & -4.556951     & 0.0401 7188 870 & 0.0016 1378 \\
\hline
7 $^1$S & 548 & 1.518309 & -4.3430 5141 3142 &               & 0.0969 7843 517 & 0.0094 0482 \\
7 $^1$S & 631 & 1.518309 & -4.3834 3400 6079 &               & 0.0913 8755 388 & 0.0083 5168 \\
7 $^1$S & 722 & 1.307501 & -4.4522 1155 3712 &               & 0.0798 8670 608 & 0.0063 8189 \\
7 $^1$S & 820 & 0.885885 & -4.5295 4113 3144 &               & 0.0496 8399 115 & 0.0024 6850 \\
7 $^1$S & 820 & 0.655102 & -4.5408 7695 5351 & -4.541692     & 0.0328 7934 851 & 0.0010 8105 \\
\hline
8 $^1$S & 631 & 1.610647 & -4.1604 9770 5052 &  & 0.0998 0414 110 & 0.0099 6087 \\
8 $^1$S & 722 & 1.399839 & -4.2926 5026 8268 &  & 0.0921 0938 521 & 0.0084 8414 \\
8 $^1$S & 792 & 1.307501 & -4.4056 9055 9470 &  & 0.0778 3110 840 & 0.0060 5768 \\
8 $^1$S & 820 & 0.885885 & -4.4896 9332 2758 &  & 0.0595 8906 504 & 0.0035 5086 \\
8 $^1$S & 820 & 0.564485 & -4.5285 0740 1021 &  & 0.0432 0219 907 & 0.0018 6643 \\
\hline\hline
\end{tabular}}
\end{center}
\end{table}


\begin{table}
\caption{Transition amplitudes and final state probabilities for the nuclear
$\beta^{-}$ decay from the ground $^2S$-state of Li atom into the ground and
excited $n^2S$-states ($n = 1, \ldots , 5$) of the Be$^+$ ion.}
\begin{center}
\scalebox{0.83}{
\begin{tabular}{crccclc}
\hline\hline
State of Be$^+$ & N $\qquad$  & $\qquad$  Amplitude $\qquad$  & $\qquad$  Probability $\qquad$  &  $\qquad$ Energy (a.u.) $\qquad$   & Ref. Ener.$^{a,b}$ & Ref.  \\
\hline
  1 $^2$S   & 33   & 0.7663 5321 7067 & 0.5872 9725 34 & -14.3101 7148 9890 & -14.3247 6317 6790 43(22) & \cite{Pachucki}\\
  1 $^2$S   & 98   & 0.7471 4678 5069 & 0.5582 2831 85 & -14.3180 5814 4763 &             &  \\
  1 $^2$S   & 216  & 0.7589 3103 1118 & 0.5759 7631 00 & -14.3211 8401 2888 &             &   \\
  1 $^2$S   & 403  & 0.7577 2587 1800 & 0.5741 4849 68 & -14.3222 1554 7857 &             &   \\
\hline
  2 $^2$S   & 98   & 0.5252 0691 4085 & 0.2758 4230 26 & -13.9165 4421 8865 & -13.9227 8926 85442 & \cite{Pachucki2}\\
  2 $^2$S   & 216  & 0.5219 0688 5130 & 0.2723 8679 67 & -13.9182 6103 3217 &             &   \\
  2 $^2$S   & 403  & 0.5224 7452 7076 & 0.2729 7963 15 & -13.9191 9775 7138 &             &    \\
\hline
  3 $^2$S   & 98   & 0.0948 3616 3164 & 0.0089 9389 78 & -13.7907 5653 8607 & -13.7987 1657 & \cite{King}\\
  3 $^2$S   & 216  & 0.0788 4013 8714 & 0.0062 1576 75 & -13.7952 3708 6511 &               &  \\
  3 $^2$S   & 403  & 0.0879 8494 1014 & 0.0077 4134 98 & -13.7957 6122 3749 &               &  \\
\hline
  4 $^2$S   & 216  & 0.0470 1746 7961 & 0.0022 1064 23 & -13.7255 3023 7584 & -13.7446 306 & \cite{Wang}   \\
  4 $^2$S   & 403  & 0.0335 5753 9656 & 0.0011 2610 85 & -13.7352 7047 2759 &                 & \\
\hline
  5 $^2$S   & 403  & 0.0177 1728 0830 & 0.0003 1390 20 & -13.7015 2700 4998 & -13.7162 8624 & \cite{King}\\
\hline\hline
\end{tabular}}
\end{center}
\footnotetext[1]{The reference energies of the here employed Li ground state wave functions are
$-7.470554439852$, $-7.472999997446$, $-7.475065436133$ and $-7.475819455001$ a.u. for $n=3,4,5,6$ wave functions, respectively.
Best ground state energy for neutral Li is $-7.47806032391010(32)$ a.u. \cite{Pachucki}.}
\end{table}

\newpage


\begin{table}
\caption{Transition amplitude and final state probability for the nuclear
$\beta^+$ decay from the ground $1^1S$-state of the Be atom into the
ground $1^1S$-state of the Li$^-$ ion.}
\begin{center}
\begin{tabular}{rcccc}
\hline\hline
    N   &  $\qquad$ Amplitude $\qquad$    & $\qquad$  Probability $\qquad$ &  $\qquad$  Energy (a.u.) $\qquad$ &  $\qquad$ Ref. Ener. (a.u.)   \\
\hline
    572  & 0.4651 9096 9919 & 0.2164 0263 85 & -7.4970 6618 7901 & -7.5005 82500$^a$ \\
    1001 & 0.4539 8563 2775 & 0.2061 0295 48 & -7.4985 2137 5044 &              \\
    2155 & 0.4544 2664 7590 & 0.2065 0357 80 & -7.4989 1384 5101 &              \\
\hline\hline
\end{tabular}
\end{center}
\footnotetext[1]{Ref. \cite{Frolov-Li-}.}
\end{table}

\newpage


\begin{table}
\caption{Transition amplitudes and final state probabilities for the nuclear
$\beta^{+}$ decay from the ground and excited $n^1S$-states ($n$ = 1, 2,
3) of the Be atom into the ground $1^1S$-state of the Li$^-$ ion.}
\begin{center}
\begin{tabular}{crcccc}
\hline\hline
  State of Be     &  N  & $\qquad$ Amplitude $\qquad$ & $\qquad$ Probability $\quad$ & $\quad$ Energy (a.u.) $\quad$ &  Ref. Ener.$^a$ (a.u.) \\
\hline
     1 $^1$S   & 572  & 0.4651 9096 9919 & 0.2164 0263 85 & -14.6629 2117 510 & -14.6673 56486   \\
     1 $^1$S   &1001  & 0.4539 8563 2775 & 0.2061 0295 48 & -14.6648 2313 727 &                 \\
     1 $^1$S   &2155  & 0.4544 2664 7590 & 0.2065 0357 80 & -14.6651 8511 672 &                 \\
\hline
     2 $^1$S   & 572  & 0.6053 8860 5257 & 0.3664 9536 34 & -14.4073 4275 418 & -14.4182 40328   \\
     2 $^1$S   &1001  & 0.5941 2364 3811 & 0.3529 8290 41 & -14.4141 1323 581 &                 \\
     2 $^1$S   &2155  & 0.5959 8607 9773 & 0.3551 9940 73 & -14.4152 5746 639 &                 \\
\hline
     3 $^1$S   & 572  & 0.0000 5704 4101 & 0.0000 0000 32 & -14.3594 8753 410 & -14.3700 87876   \\
     3 $^1$S   &1001  & 0.0000 8923 5928 & 0.0000 0000 80 & -14.3616 2061 147 &                 \\
     3 $^1$S   &2155  & 0.0000 9848 3872 & 0.0000 0000 97 & -14.3617 8613 448 &                 \\
\hline\hline
\end{tabular}
\end{center}
\footnotetext[1]{The reference energies of the here employed Li$^-$ ground state wave functions are
$-7.4970 6618 7901$ a.u. for 572, $-7.4985 2137 5044$ a.u. for 1001, and
$-7.4989 1384 5101$ a.u. for 2155 configurations, respectively. Best energy is $-7.5005 82500$ a.u. \cite{Frolov-Li-}.}
\footnotetext[2]{{\footnotesize Ref. \cite{Adam-Be}.}}
\end{table}

\newpage


\begin{table}
\caption{The transition amplitudes and final state probabilities for the
nuclear $\beta^-$ decay from the ground $1^1S$-state of the Be atom
into the ground $1^1S$-state of the B$^+$ ion. Convergence of the
calculations is shown with the increasing basis set.}
\begin{center}
\begin{tabular}{crllll}
\hline\hline
  State of B$^+$  & N   & $\qquad$  Amplitude $\qquad$  & $\qquad$ Probability $\qquad$  & $\qquad$  Energy (a.u.) $\qquad$  & Ref. Ener.    \\
\hline
  1 $^1$S   & 572  & 0.751759184759 & 0.5651418719 & -24.344002956525 & -24.348884446$^a$  \\
  1 $^1$S   &1001  & 0.749760917799 & 0.5621414339 & -24.345883907324 &                 \\
\hline\hline
\end{tabular}
\end{center}
\footnotetext[1]{Ref. \cite{Komasa-B+}.}
\end{table}

\end{document}